\documentclass[]{aiaa-tc}
\usepackage[T1]{fontenc}
\usepackage[latin9]{inputenc}
\usepackage{geometry}
\usepackage{color}
\usepackage{booktabs}

\usepackage{amsmath}
\usepackage{amssymb}
\usepackage{stmaryrd}
\usepackage{graphicx}
\usepackage{setspace}
\usepackage{esint}
\usepackage{multirow}

 \usepackage{varioref}
 \usepackage{wrapfig}
 \usepackage{threeparttable}
 \usepackage{dcolumn}
  \newcolumntype{d}{D{.}{.}{-1}}
 \usepackage{nomencl}
  \makeglossary
 \usepackage{subfigure}
 \usepackage{subfigmat}
 \usepackage{fancyvrb}
  \fvset{fontsize=\footnotesize,xleftmargin=2em}
 \usepackage{lettrine}
 \usepackage[dvips]{dropping}

\usepackage{color}
  \definecolor{darkblue}{rgb}{0.0,0.0,0.5}
  \definecolor{lightgray}{rgb}{0.95,0.95,0.95}
  \definecolor{darkred}{rgb}{0.8,0.0,0} 

\newcommand{\mr}[1]{\mathrm{#1}}

\newcommand{\f}[1]{\overline{#1}} 
\newcommand{\ff}[1]{\widetilde{#1}} 
\newcommand{\vect}[1]{\boldsymbol{\mr{#1}}}
\newcommand{\tens}[1]{\boldsymbol{\mathsf{#1}}}

\newcommand{\pd}[2]{\frac{\partial{#1}}{\partial{#2}}} 
\newcommand{\tpd}[2]{\partial{#1}/\partial{#2}} 

 \title{Coherent-vorticity Preserving Large-Eddy Simulation of trefoil knotted vortices}

 \author{
  Zongxin Yu$^1$, Jean-Baptiste Chapelier$^1$
  \ and Carlo Scalo$^{1,}$ \thanks{Corresponding author, scalo@purdue.edu}\\
  {\normalsize\itshape
    $^1$School of Mechanical Engineering and Aeronautical Engineering, Purdue University, IN, USA}\\
  }

 \AIAApapernumber{YEAR-NUMBER}
 \AIAAconference{AIAA SciTech Forum and Exposition, January 2018, Florida}
 \AIAAcopyright{\AIAAcopyrightD{YEAR}}

\begin{document}

\maketitle

\begin{abstract} 
We have performed Coherent-vorticity Preserving Large-Eddy simulations of a trefoil knot-shaped vortex, inspired by the experiments of Kleckner and Irvine~\cite{kleckner2013creation}.
The flow parameter space is extended in the present study, including variations of the circulation Reynolds numbers in the range $Re_\Gamma = 2\times 10^3 - 200\times 10^3$, where $Re_\Gamma =20,000$ is the value used in the experiments.
The vortex line corresponding to the trefoil knot is defined using a parametric equation and the Biot-Savart law is employed to initialize the velocity field. 
The CvP LES computation displays a good qualitative match with the experiment.
In particular, the vortex entanglement process is accurately represented as well as the subsequent separation of the main vortex in two distinct structures - a small and a large vortex - with different self-advection speeds that have been quantified.
The small vortex propagates faster than  the large oscillatory vortex which carries an important amount of vorticity. 
The advection velocity of the vortex before bursting is found to be independent of the Reynolds number.
The low Reynolds number computation leads to a decrease of the separated vortices velocity after bursting, compared to the higher Reynolds computations.
The computation of energy spectra emphasizes intense energy transfers from large to small scales during the bursting process.
The evolution of volume-averaged enstrophy shows that the bursting leads to the creation of small scales that are sustained a long time in the flow, when a sufficiently large Reynolds number is considered ($Re_\Gamma>20000$).
The low Reynolds number case $Re_\Gamma=2,000$ hinders the generation of small scales during the bursting process and yields essentially laminar dynamics.
The onset of background turbulence due to the entanglement process can be observed at $Re_\Gamma = 200,000$.

\end{abstract}

\section{Introduction}
The study of entangled vortex filaments is of interest since most of turbulent flows show the presence of such phenomena, as seen in Direct Numerical Simulations by She et al.~\cite{she1990intermittent}, where long-lived high-amplitude vortex are reported as tube-like structures which are a key component of turbulent flows. In 1991, Vincent and Meneguzzi~\cite{vincent1991spatial} further confirmed that the width of these tubes is of the order of a few dissipation scales, while their length can reach the integral scale of the flow. Kerr~ \cite{kerr1985higher} studied third- and fourth-order correlations and reported strong alignment between the vorticity, rate of strain, and scalar-gradient fields with DNS. The study on equilibrium turbulent flow fields by Jimenez et al.~\cite{JimenezWSR_JFM_1993} showed that self-stretching is not important in the evolution of vortices. 

In the present paper, we propose a setup to study numerically the dynamics of knotted vortices which provide a solid framework for the fundamental study of entangled vortices.
We aim in particular at reproducing numerically a recently published experimental study by Kleckner and Irvine~\cite{kleckner2013creation}.
A knotted vortex path is defined using a parametric equation, and the corresponding velocity field is found using the Bio-Savart law, coupled with a smoothing kernel defining the vortex core size.
A recently developed LES model~\cite{ChapelierSW_AIAA_2017} which is able to capture accurately the dynamics of transitional vortical flows with marginal resolution is considered.

The outline of the paper is as follows: Section III.\ref{sec:validation} shows a qualitative validation of the present simulation with the experiment.
Section II introduces the governing equations and the LES model considered in the present study.
Section III.A presents the initialization and simulation parameters relevant to the trefoil-knot vortex flow.
Section III.B features a validation of the numerical setup with experiments.
Section III.\ref{sec:statistics} is devoted to the physical study of the dynamics of the trefoil knot vortex and Section III.\ref{sec:kinematics} focuses on flow kinematics.

\section{Methodology}

\subsection{Governing equations}
The flow motion considered in the present study is assumed to be governed by the set of compressible Navier-Stokes equations:
\begin{equation}
\mathcal{NS}(\vect{w})=\pd{\vect{w}}{t}+\vect{\nabla}\cdot\left[\tens{F}_{\mr{c}}(\vect{w})-\tens{F}_{\mr{v}}(\vect{w},\nabla\vect{w})\right]=\vect{0},
\end{equation}
where $\vect{w}=\left(\rho,\rho\vect{U},\rho E\right)^{\mr{T}}$ is the vector of conserved variables $\rho$, $\vect{U}$ and $E$, density, velocity and total energy respectively, and $(\nabla\vect{w})_{ij} = \tpd{w_i}{x_j}$ its gradient.
The viscous and convective flux tensors $\tens{F}_{\mr{c}},\tens{F}_{\mr{v}}\in\mathbb{R}^{5\times3}$ read 
\begin{equation}
\tens{F}_{\mr{c}} =
\begin{pmatrix}
\rho\vect{U}^{\mr{T}}\\
\rho\vect{U}\otimes\vect{U}+ p\tens{I}\\
(\rho E+p )\vect{U}^{\mr{T}}
\end{pmatrix},\quad\text{and}\quad 
\tens{F}_{\mr{v}} = 
\begin{pmatrix}
\vect{0}\\
\tens{\tau}\\
\tens{\tau}\cdot\vect{U}-\lambda\vect{\nabla} T^{\mr{T}}
\end{pmatrix},
\end{equation}
where $T$ is the temperature, $p$ is the pressure, $\lambda$ is the thermal conductivity of the fluid and $\tens{I} \in \mathbb{R}^{3\times3}$ is the identity matrix.
For a Newtonian fluid, we have 
\begin{equation}
\tens{\tau}=2\mu\tens{S},
\end{equation}
where $\mu$ is the dynamic viscosity and 
\begin{equation}
\tens{S}=\frac{1}{2}\left[\nabla\vect{U}+\nabla\vect{U}^{\mr{T}}-\frac{2}{3}\left(\vect{\nabla}\cdot\vect{U}\right)\tens{I}\right].
\label{eq:shear:stress}
\end{equation}
The ideal gas law is considered for the closure of the system of equations, namely,
\begin{equation}
p=(\gamma-1)\left(\rho E-\frac{1}{2}\rho\vect{U}\cdot\vect{U}\right),
\end{equation}
where $\gamma$ is the heat capacity ratio.

The LES equations are obtained by applying a low-pass filter to the Navier-Stokes equations~\cite{leonard1974energy}.
The spatial filtering operator applied to a generic quantity $\phi$ reads 
\begin{equation}
\f{\phi}(\vect{x},t)=g (\vect{x})\star\phi (\vect{x}),
\label{eq:filt}
\end{equation}
where $\star$ is the convolution product and $g\left(\vect{x}\right)$ is a filter kernel related to a cutoff length scale $\overline{\Delta}$ in physical space~\cite{sagautLESbook}. The compressible case requires density-weighted filtering approaches. The density-weighted or Favre filtering operator is defined as
\begin{equation}
\ff{\phi}=\frac{\f{\rho\phi}}{\f{\rho}}.
\end{equation}

In the present study, the compressible LES formalism introduced by Lesieur et al.~\cite{lesieur2001favre,lesieur2005large,lesieur1996new} is adopted yielding the following set of filtered compressible Navier-Stokes equations:
\begin{equation}
\mathcal{NS}(\f{\vect{w}}) = \vect{\nabla} \cdot \tens{F}_{\mr{SGS}}(\f{\vect{w}},\nabla\f{\vect{w}}),
\label{eq:filtns}
\end{equation}
where $\f{\vect{w}}=\left(\f{\rho},\f{\rho}\ff{\vect{U}},\f{\rho}\ff{E}\right)^{\mr{T}}$ is the vector of filtered conservative variables.

The SGS tensor $\tens{F}_{\mr{SGS}}$ is the result of the filtering operation and it encapsulates the dynamics of the unresolved sub-grid scales, and is modeled here using the eddy-viscosity assumption: 
\begin{equation}
\tens{F}_{\mr{SGS}}(\f{\vect{w}},\nabla\f{\vect{w}}) = 
\begin{pmatrix}
\vect{0}\\
2\mu_{t}\f{\tens{S}}\\
-\frac{\mu_{t}C_{p}}{Pr_{t}}\vect{\nabla}\ff{T}^{\mr{T}}
\end{pmatrix},
\end{equation}
where $\f{\tens{S}}$ is the shear stress tensor computed from equation~\eqref{eq:shear:stress} based on the Favre-filtered velocity $\ff{\vect{U}}$, $Pr_{t}$ is the turbulent Prandtl number, which is set to 0.5~\cite{erlebacher1992toward}, $C_{p}$ is the heat capacity at constant pressure of the fluid and $\mu_{t}$ is the eddy-viscosity which depends on the chosen sub-grid model.

In the present work, the CvP-Smagorinsky closure is adopted, which yields accurate results for transitional and high-Reynolds number flows~\cite{ChapelierSW_AIAA_2017}.
The eddy viscosity for this closure reads:
\begin{equation}
\mu_t=\rho f(\sigma) (C_{\mathrm{S}}\overline{\Delta})^2\sqrt{S_{ij}S_{ij}}
\end{equation}
where $C_{\mathrm{S}}=0.172$ is the Smagorinsky constant, $\overline{\Delta}$ is the grid size and $f(\sigma)$ is the CvP turbulence sensor built from the ratio $\sigma$ of test-filtered to grid-filtered enstrophy. This sensor attenuates the intensity of SGS dissipation of transitional turbulence or coherent vortices. 

\subsection{Numerical method}
The compressible, Favre-filtered Navier-Stokes equations are solved using a 6th order compact finite difference scheme solver originally written by Nagarajan \emph{et al.}~\cite{nagarajan2003robust}, currently under development at Purdue University.
The solver is based on a staggered grid arrangement, providing superior accuracy compared to a fully collocated approach~\cite{lele1992compact}.
The speed of sound is adjusted in such a way that the local Mach number is not above 0.1 for the computations performed in the present study, in order to recover almost incompressible flow dynamics.
The time integration is performed using a third order Runge-Kutta scheme.

\subsection{Flow initialization using the Biot-Savart law}
\begin{figure}
\centering
\includegraphics[width=.8\linewidth]{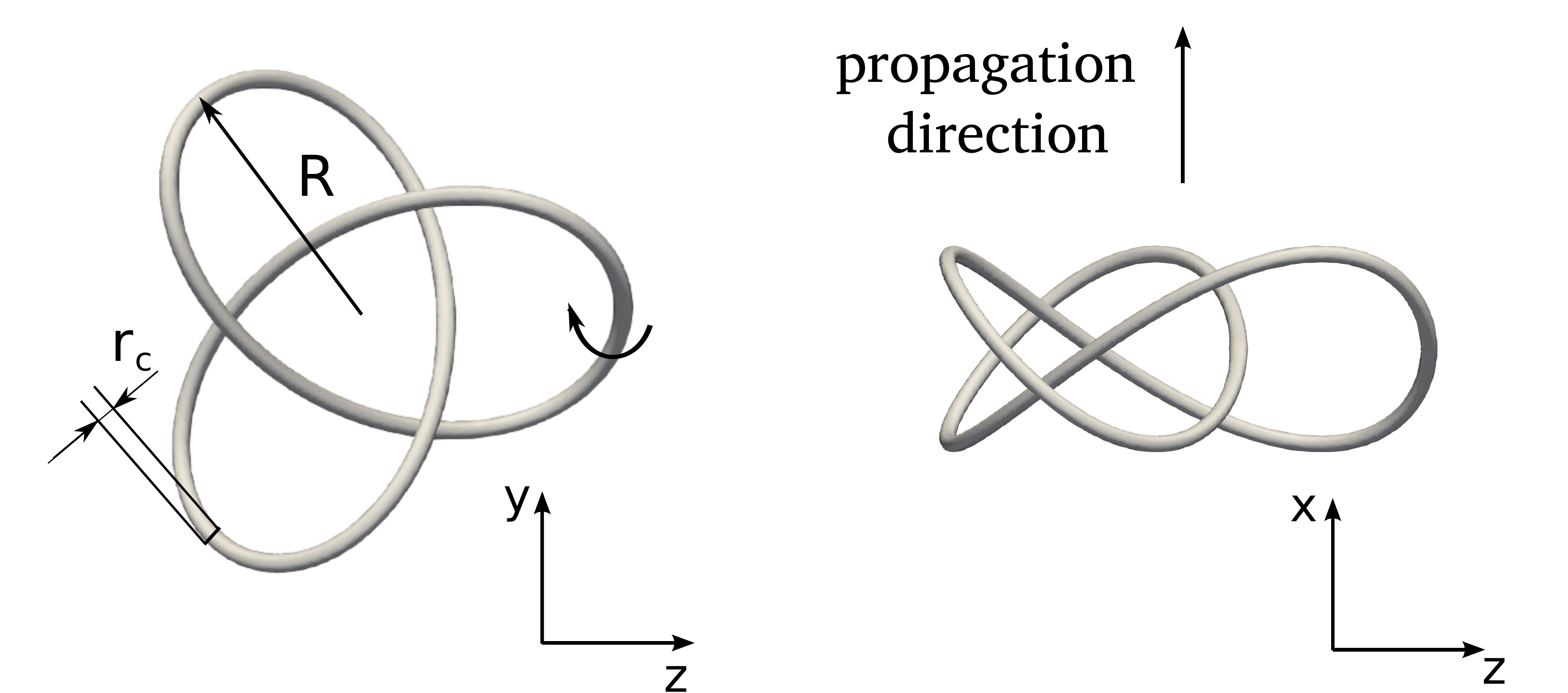}

\caption{Initial vortex configuration defined by parametric equation.}
\label{fig:setup}
\end{figure}

In this section, a fundamental test case is defined for the numerical study of knotted vortices.
A vortex filament corresponding to a trefoil knot is initialized in a periodic, cubic box.
The filament equation reads:
\begin{equation}
\bold{X}(\theta) = \left[-R_0\sin(3\theta), R_0\left(\sin(\theta)+2\sin(2\theta)\right),R_0 \left(\cos(\theta)-2\cos(2\theta)\right)\right]
\end{equation}
where $R_0=R/3$ and is $R$ knot radius.
The velocity field induced by the vortex filament is determined by the Biot-Savart law.
\begin{equation} \label{eq:biot_savart}
\bold{u}(\bold{x})=-\frac{\Gamma}{4\pi}\int{K_v\frac{\left(\bold{x}-\bold{X}(\theta)\right)\times\bold{t}(\theta)}{\left|\bold{x}-\bold{X}(\theta)\right|^3}d\theta}
\end{equation}
where $\bold{t}(\theta)$ is the tangent vector to the helical filament, $\Gamma$ is the circulation and $K_v$ is a function allowing to define the shape of the vortex core~\cite{bagai1993flow} and reads: 
\begin{equation} \label{eq:smoothing_kernel}
K_v=\frac{\left|\bold{x}-\bold{X}(\theta)\right|^3}{\left(\left|\bold{x}-\bold{X}(\theta)\right|^{2n}+r_c^{2n}\right)^{\frac{3}{2n}}}
\end{equation}
where $r_c$ is the core radius.
The case $n=\infty$ corresponds to a Rankine vortex.
The value of $n=4$  is adopted to achieve a smooth transition between the inner, rotational flow and the outer, potential flow.
The initial condition and the relevant geometrical parameters for the knottex vortex study are presented in figure~\ref{fig:setup}.

\section{Simulation of a trefoil knot vortex at high Reynolds number}
\subsection{Simulation parameters} \label{sec:parameters}
In this section, computations performed using the present numerical methodology are compared against experiments in order to demonstrate its relevance and accuracy for the problem considered.

Simulations are performed in a cubic computational domain $\Omega=[0,10R_0]^3$. The trefoil knot vortex is initialized at the center of the domain. All simulations are run with circulation  $\Gamma=0.02$ and   $r_c/R=0.088$. Details are shown in Table~\ref{tab:case}. These parameters correspond approximately to the experimental study of Kleckner et al.~\cite{kleckner2013creation}, where $Re_{\Gamma}\sim 10000$ and $r_c/R \sim 0.05$. 
The vortex core parameter $rc/R=0.088$ is selected in this computation to define a sufficient number of grid points inside the vortex core to yield an accurate representation of the dynamic of the knotted vortex and subsequent bursting.
The LES computation features $192^3$ grid points leading to the definition of 10 points inside the vortex core.

The LES model considered is the CvP approach~\cite{ChapelierSW_AIAA_2017}, which has been found to be successful for the prediction of transitional flows using coarse grids.
A non-dimensional time $t^*=t\Gamma/R^2$ is used to display transient data. 

\begin{table}
\centering
\begin{tabular}{p{2cm}<{\centering} p{2cm}<{\centering} p{2cm}<{\centering} p{2cm}<{\centering}} 
\toprule
 Case & Re & $r_c/R$ & Grid Size\\
\midrule                                                                                                                                                                                                                                                                                                                                                                                                                                                                                                                                                                                                                                                                                                                                                                                                                                                                                                                         
(a) & 2,000 & 0.088 & $192^3$ \\
(b) & 20,000 & 0.088 & $192^3$ \\
(c) & 200,000 & 0.088 & $192^3$ \\
\bottomrule 
\end{tabular} 
\caption{Vortex parameter set up}
\label{tab:case}
\end{table}
\subsection{Comparison against experiments} \label{sec:validation}
\begin{figure}
\centering
\includegraphics[width=.8\linewidth]{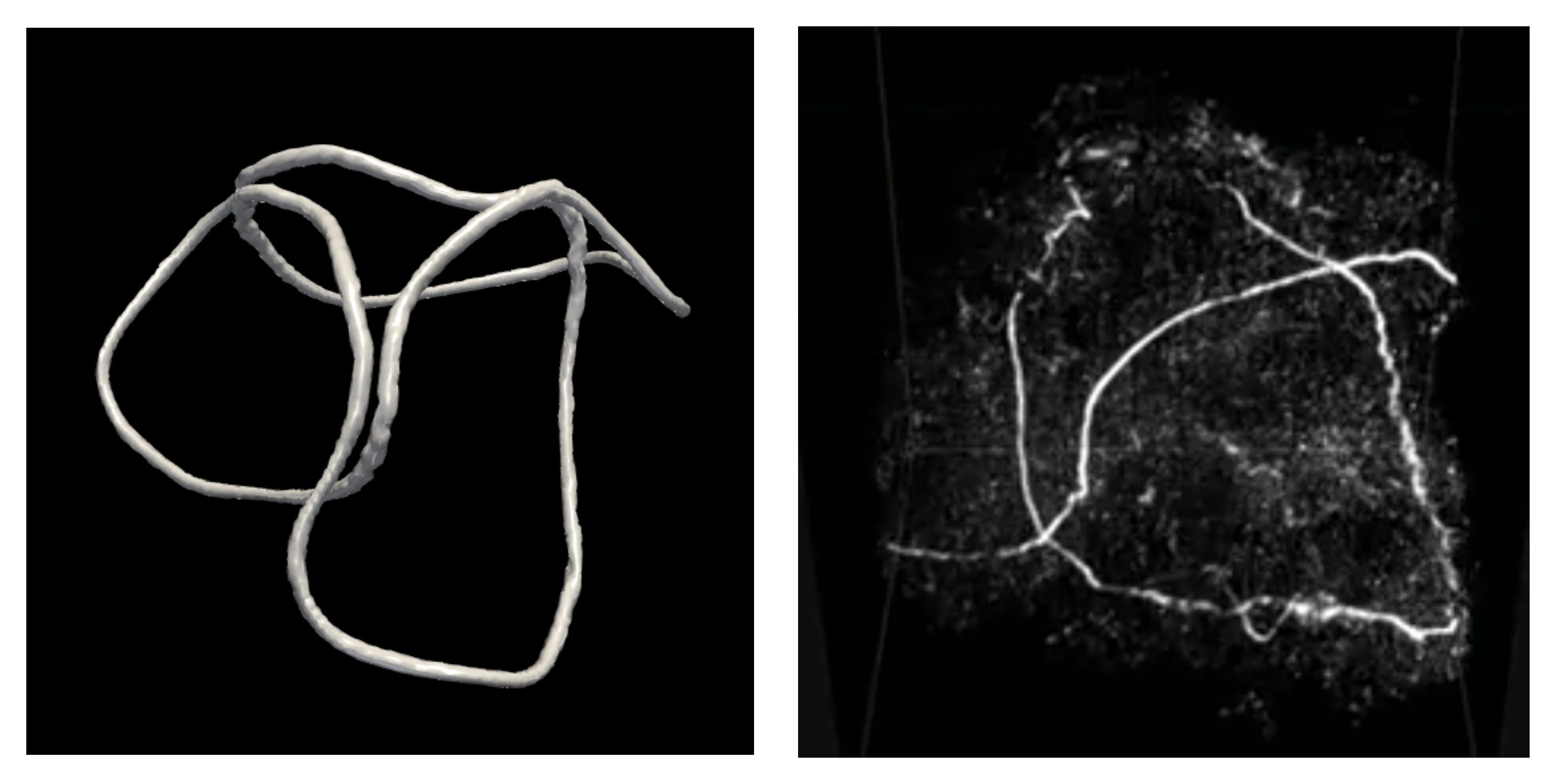}\\
\vspace*{0.4cm}
\includegraphics[width=.8\linewidth]{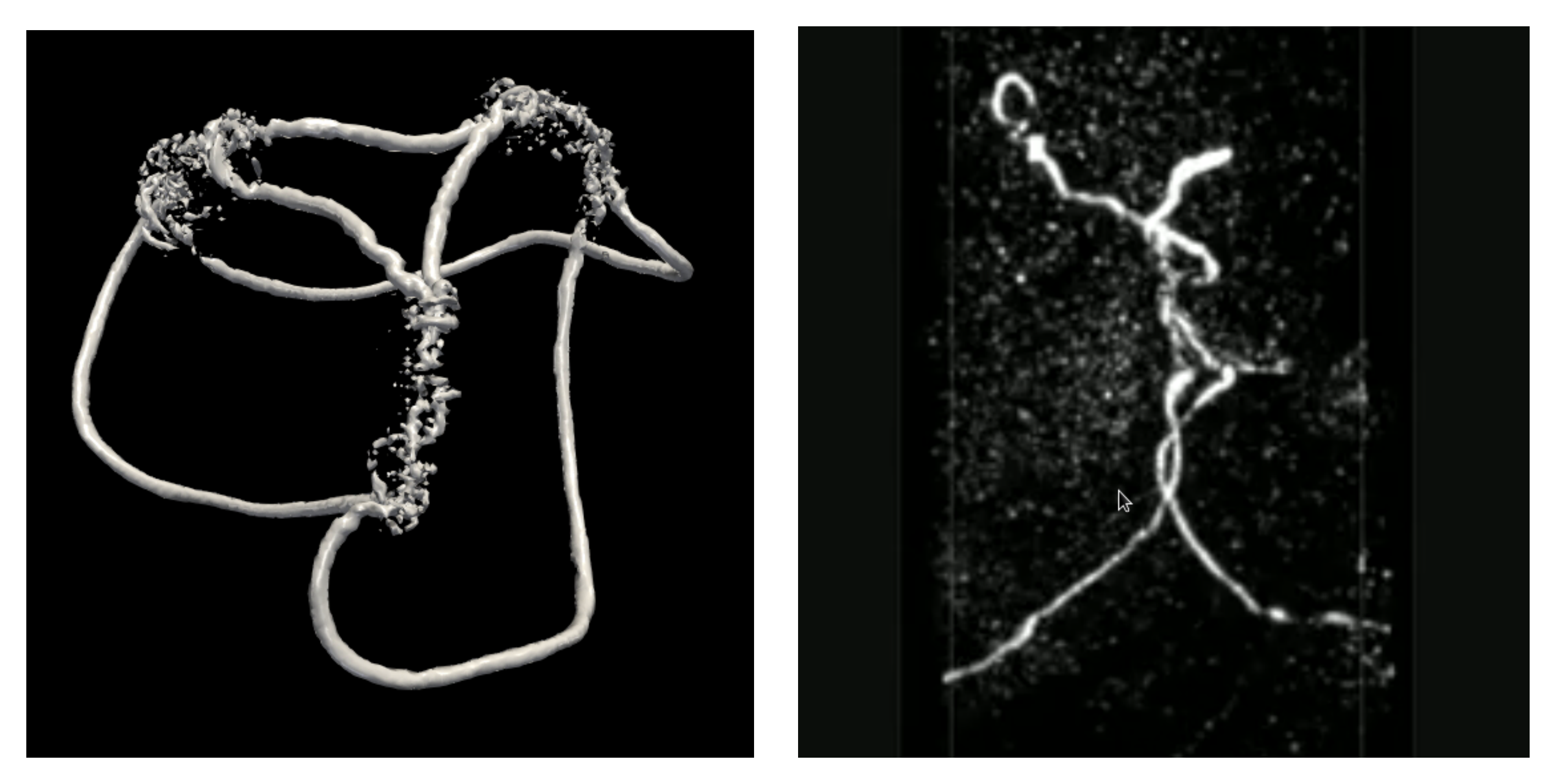}\\
\vspace*{0.4cm}
\includegraphics[width=.8\linewidth]{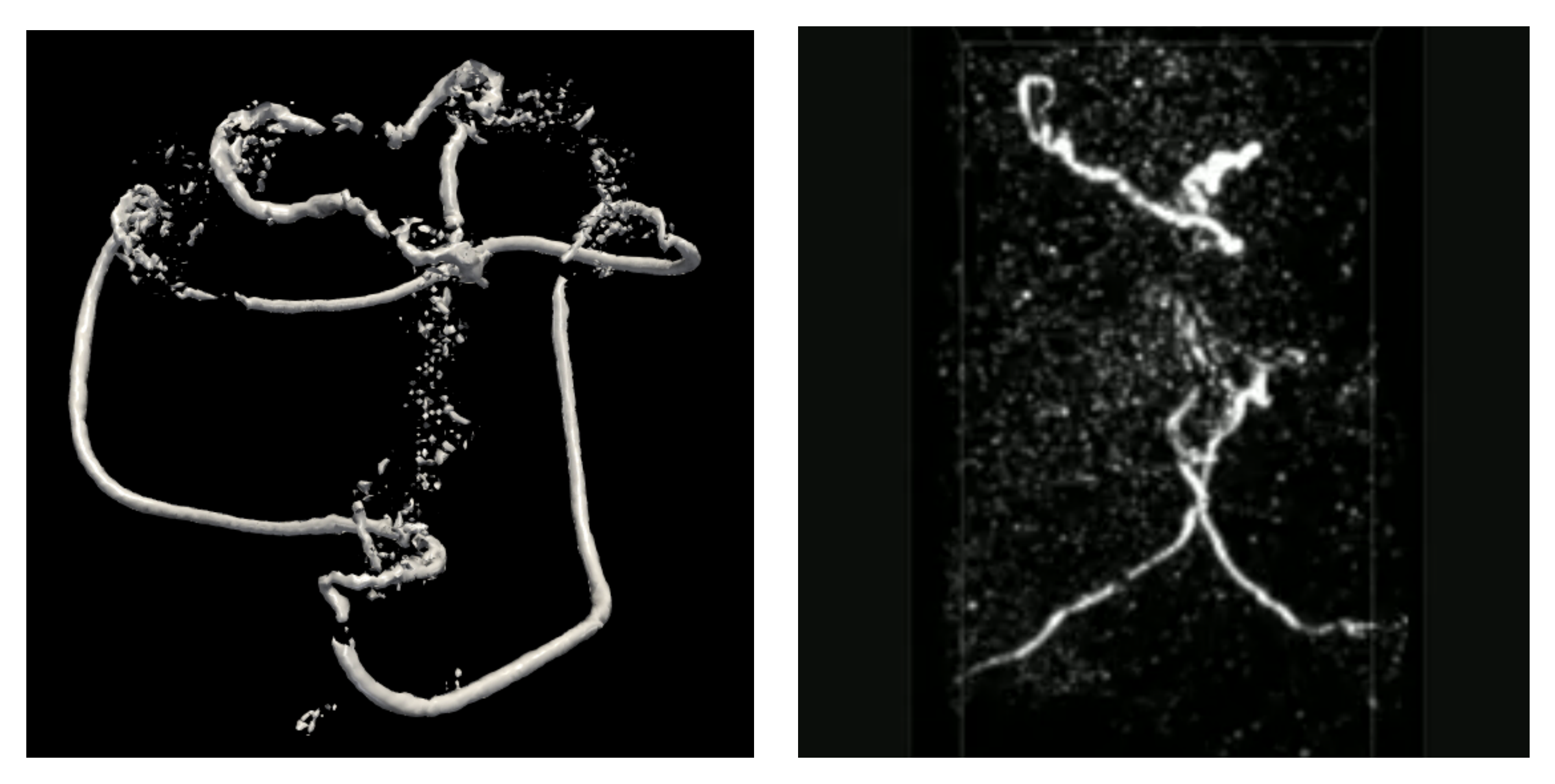}
\caption{Comparison of flow topology between CvP-LES computation (left) and experiment of Kleckner and Irvine~\cite{kleckner2013creation} at three different times(from top to bottom): before, during and after the vortex entanglement process.}
\label{fig:compexp}
\end{figure}

Figure~\ref{fig:compexp} shows the flow topology extracted from the simulation (b)(left), and compared to the experimental results (right) with same  circulation Reynolds number $Re_\Gamma$ and knotted vortex radius $R$.
Three different times are considered for the comparison.
The top row shows the flow before the vortex filament connection, when the initial trefoil knot vortex structure is still intact and begins to stretch.
The two plots compare well, which means that the inviscid dynamics of the vortex evolution are well characterized by the simulation.
The middle row corresponds to the time of vortex-line entanglement.
The simulation is found to provide a good match with the experiment, even though all the details of the entanglement process are not captured due to the coarse resolution considered.
The last plots present the flow topology after the knot vortex has separated in two distinct structures.
The shape of the two vortices from the simulation correspond to those from the experiment.

The flow topology is found to be essentially the same between the simulation and experiment, emphasizing the accuracy of the CvP-LES methodology coupled with high-order schemes.

\subsection{Turbulent statistics} \label{sec:statistics}

\begin{figure}
\centering
\includegraphics[width=.99\linewidth]{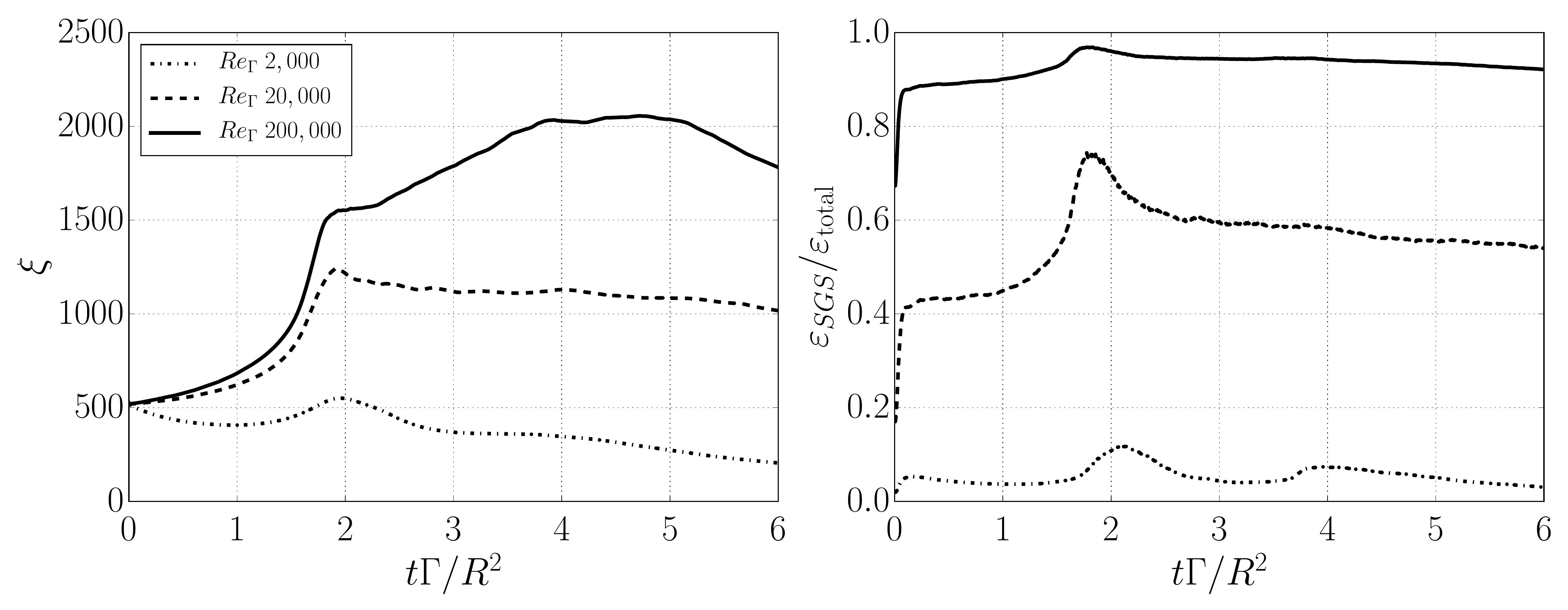}\\
\caption{Evolution of enstrophy (left) and ratio of SGS dissipation over total dissipation (right) .}
\label{fig:valenstrophy}
\end{figure}
The enstrophy is defined as $\xi = \omega\cdot\omega/2$, where $\omega$ is the vorticity vector. This quantity is sensitive to the dynamics of the small scales in the flow as these carry a significant amount of vorticity. The temporal evolution of volume-averaged enstrophy is shown in Figure~\ref{fig:valenstrophy}.

At $Re_\Gamma=2,000$, a decrease of enstrophy is observed at the beginning of the computation due to viscous effects that extract energy even from the large scales.
At $Re_\Gamma=20,000$, the enstrophy increases due to the build-up of small scales due to vortex connection and bursting, then keeps its level for a certain time, indicating the persistence of the small-scale activity after bursting, matching well with the topology in experiment as shown in Figure~\ref{fig:compexp}. The same behavior is observed at  Reynolds number 200,000, with higher levels of enstrophy caused by a more intense small-scale activity. The quick drop of enstrophy after the peak means that the small scales progressively disappear from the flow.

The relative importance of SGS dissipation compared to total dissipation increases with the Reynolds number, which means that the subgrid scale energy becomes more intense.
For the low Reynolds case $Re_\Gamma=2,000$, the SGS dissipation is very low compared to the viscous dissipation, emphasizing a marginal importance of the small scales in the flow and a DNS-like resolution.
On the other hand, in the  $Re_\Gamma=200,000$ case, the SGS dissipation levels are significantly higher than those of viscous dissipation, so the subgrid model mainly drives the dissipation in the flow, emphasizing an intense subgrid scale activity.

\begin{figure}[h!]
\centering
\includegraphics[width=.98\linewidth]{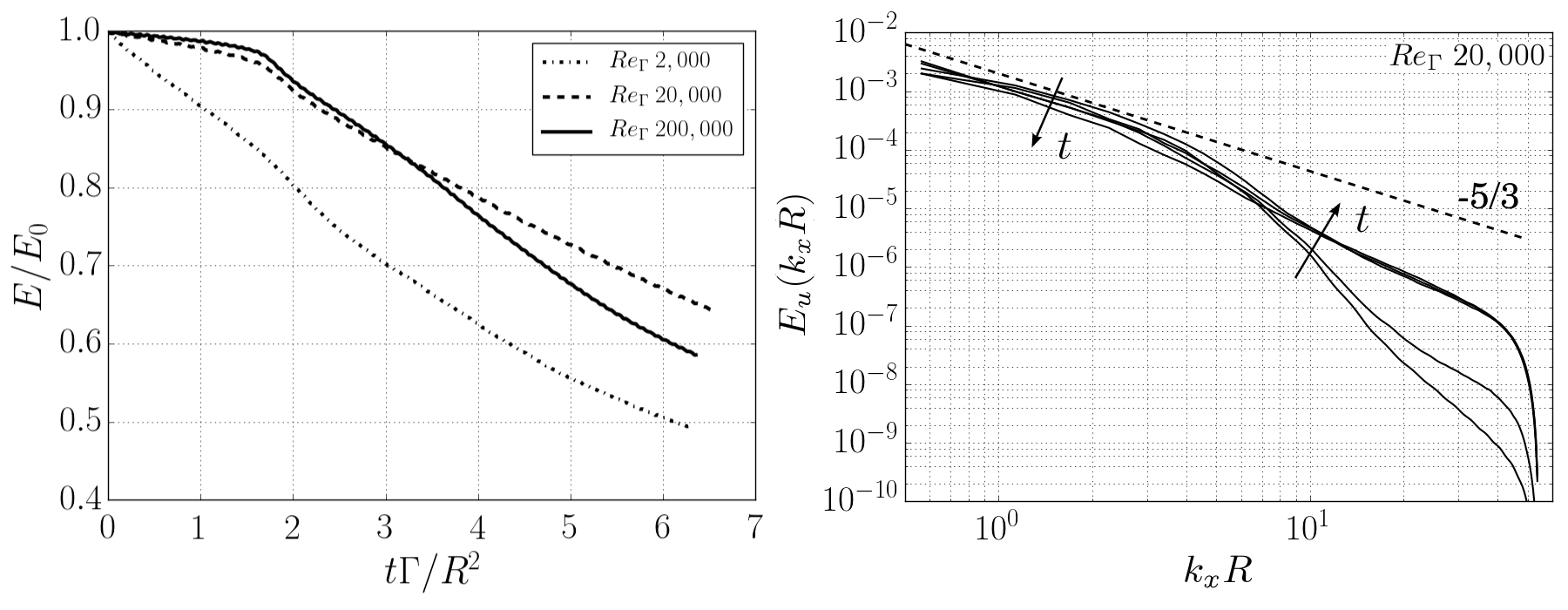}
\caption{Evolution of kinetic energy and 1D energy spectra computed in the vortex propagation direction $x$ and averaged in planes $x-z$. The arrows indicate the direction of temporal evolution ($t^*=0.73,2.12,3.50,4.89,$ and $6.28$).}
\label{LES_STATS}
\end{figure}

\begin{figure}[h!]
\centering
\includegraphics[width=.8\linewidth]{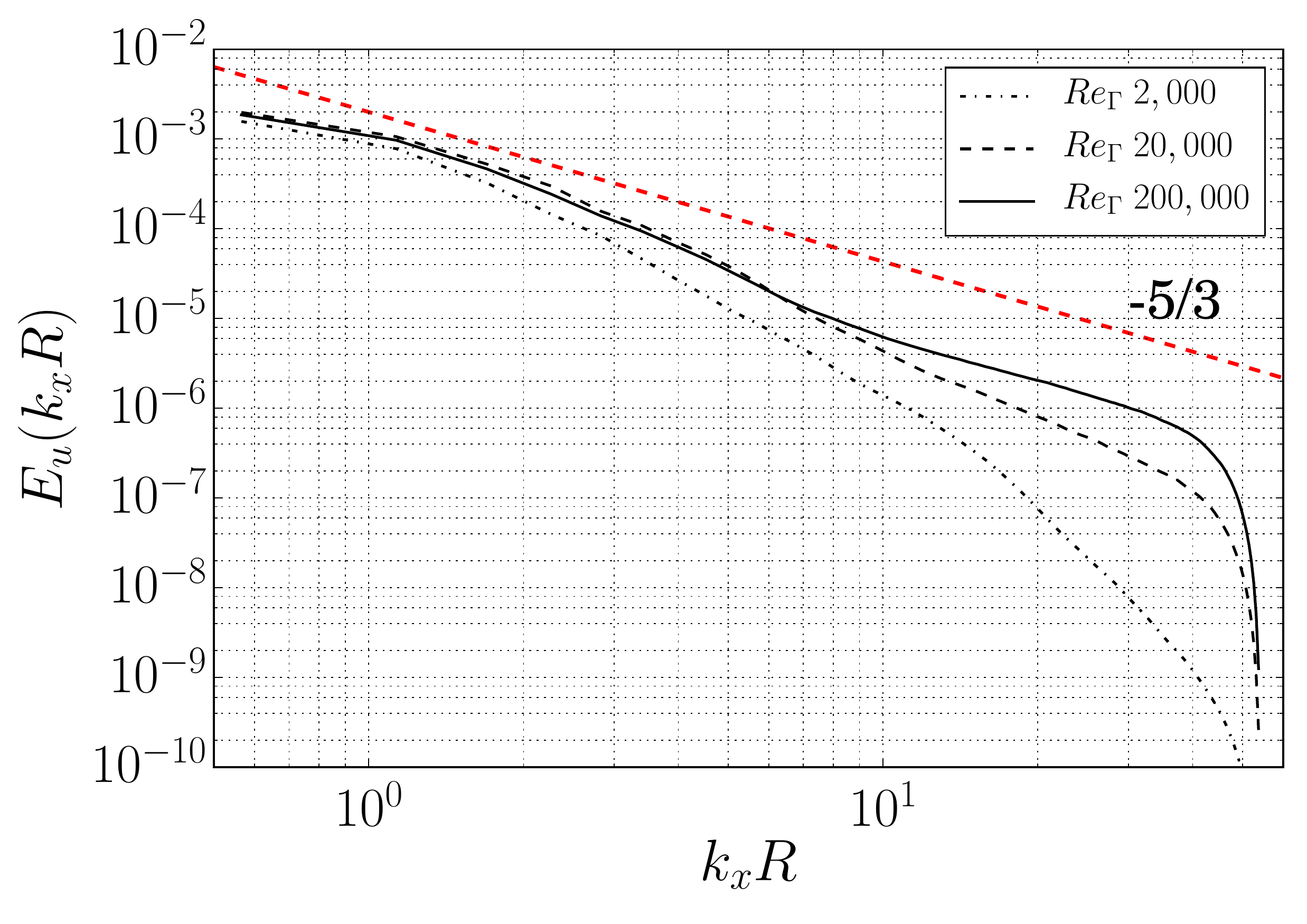}
\caption{1D energy spectrum computed at $t^\star=6$, for the various Reynolds numbers considered in the knotted vortex study.}
\label{fig:vsspectrum}
\end{figure}

As seen in figure~\ref{LES_STATS}, the volumed-averaged kinetic energy at $Re_\Gamma=2,000$ decreases monotonically due to the viscous effects in the flow.
The higher Reynolds cases $Re_\Gamma=20,000$ and $Re_\Gamma=200,000$ feature two different regimes.
At first, the evolution of large initial structures is mostly inviscid, corresponding to a conservation of kinetic energy.
The bursting process yields the creation of small scales that are strongly impacted by the viscous and subgrid dissipations, draining kinetic energy in the flow faster than during the initial regime.
The 1D kinetic energy spectrum are computed in the propagation direction and averaged in the plane y-z, at five different times before and after the bursting.
At the beginning of the simulation, it is seen that the energy is mainly concentrated in the large scales.
After the bursting at $t^*\approx 2$, a strong and sudden energy transfer from large scales to small scales is observed, as the energy spectrum steepens at high wavenumbers. 
The onset of fully developed turbulence is only observed at $Re_\Gamma=200,000$, but the enstrophy plot clearly indicates a transitional process, with the superimposition of the two well-defined coherent, triangular-shaped vortices and a background broadband turbulence.

\subsection{Breakdown and propagation kinematics} \label{sec:kinematics}

\begin{figure}[p!]
\centering
\includegraphics[width=.99\linewidth]{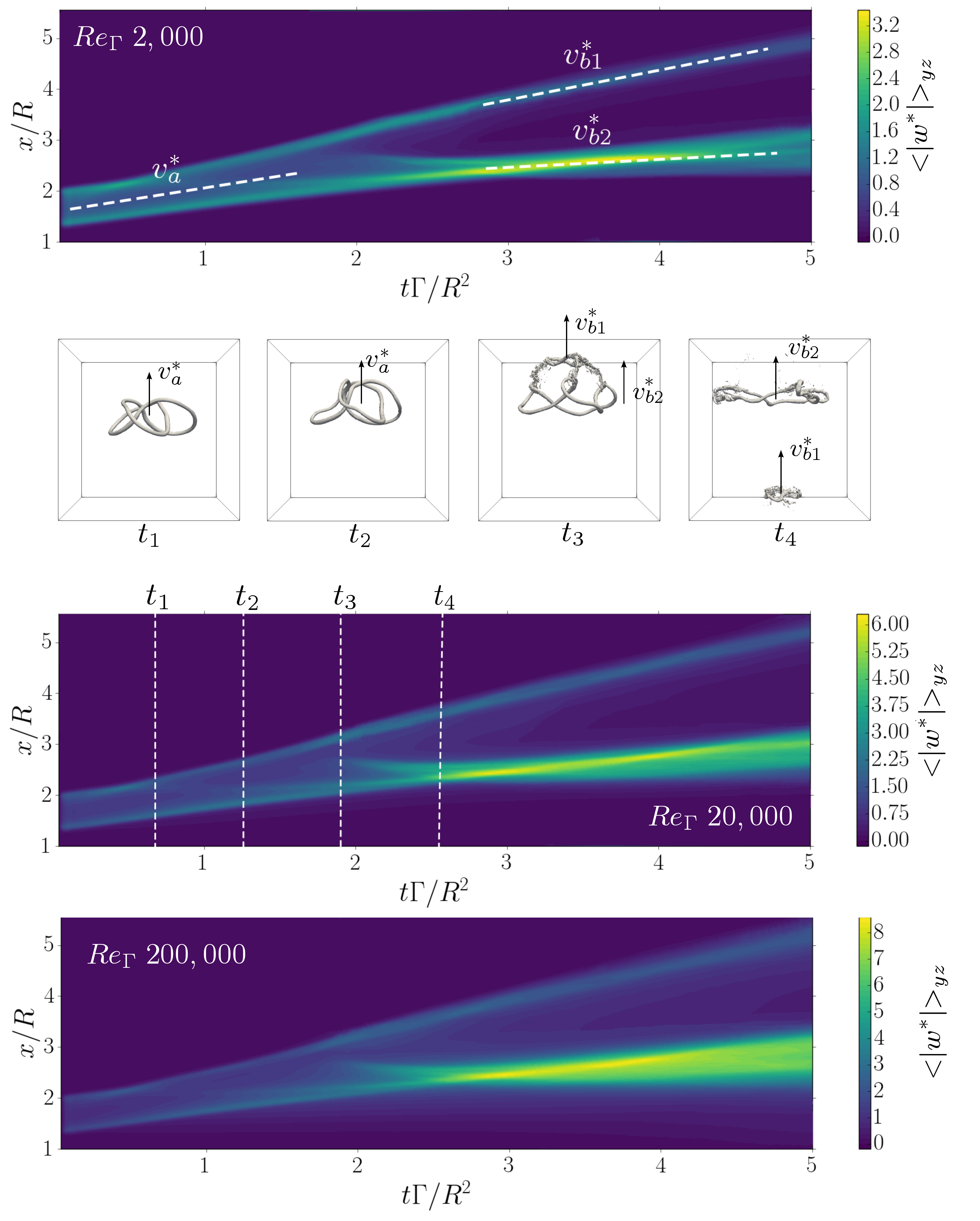}
\caption{Averaged vorticity in $y-z$ plane as a function of $x$ and $t$. From top to bottom,$Re_\Gamma=2,000,20,000$, and $200,000$ and  iso-surfaces of vorticity with $Re_\Gamma = 20,000$ at $t_1=0.63$,$t_2=1.26$,$t_3=1.89$ and $t_4=2.53$}
\label{fig:vsbranch}
\end{figure}

In this section, the flow topology is first described from visualizations of the vortices at various instants.
Figure~\ref{fig:vsbranch} shows the iso-vorticity contours for various times corresponding to the evolution of the knotted vortex.
At first, we observe the propagtion and rotation of the main structure which progressively leads to a distortion of the vortex filament.
This evolution leads at $t^*=1.26$ to the streching of the vortex and at $t^*=1.89$ to the connection of vortex filaments at three different location of the vortex.
From this merging of vortex filaments, two independent triangular vortical structures of different scales are generated.
The snapshots at $t^*=2.53$ show clearly that the two independent structures are disconnected and follow their own dynamics, as can also been seen in Figure~\ref{fig:vsbranch}.

Figure~\ref{fig:vsbranch} shows contours of the averaged modulus of vorticity in $y-z$ plane as a function of $x$ and $t$, defined as:

\begin{equation} \label{eq:avg_vor}
\begin{split}
 \omega ^*&=\omega R^2/ \Gamma \\
 \left<\omega ^*\right>_{yz}&=\frac{1}{L^2}\int_0^L \int_0^L |\omega ^*|(x,y,z,t)dy dz
\end{split}
\end{equation}

where $L$ is the box size in the respective direction of integration. This plot is interesting to characterize the path and locations of the two vortices generated by the entanglement, as the self-induction of vortices is orientied in the $x$ direction. A clear bifurcation is observed at time $t^*\approx 2$, which marks the beginning of the bursting, i.e. the separation of the knotted vortex, corresponding to the peak in Figure~\ref{fig:valenstrophy}. After the separation, the front vortex shows smaller size (Figure~\ref{fig:vsbranch}) and hence higher propagation velocity. The large vortex loop carries most of the vorticity. The highest vorticity occurs after the burst and reconnection when the large vortex loop deforms and oscillates in the y-z plane. 

For the three $Re_\Gamma$ considered, the time of bursting is similar, as seen from the branch separation in Figure~\ref{fig:vsbranch} and enstrophy peak location in Figure~\ref{fig:valenstrophy}. From Figure~\ref{fig:vsbranch}, we see that the flow kinematics are somewhat similar for the three cases before the bifurcation, i.e. independent from the Reynolds number. However, the kinematics after the bursting are affected by the Reynolds number. 
The vorticity scale in the three countours plots are different to clearly present the flow characteristics. The vorticity levels increases with the Reynolds number. Considering the highest Reynolds number, residual vorticity is detected between the separated vortex structures, corresponding to small-scale turbulent diffusion. The persistance of small scales for high Reynolds number is also confirmed by the higher enstrophy levels observed in Figure~\ref{fig:valenstrophy}. The low Reynolds number case shows a clearer separation and a smaller amplitude oscillations of the large vortex loop.  The vorticity levels are found to be higher for the large vortical structure compared to the small one, and this difference increases with the Reynolds number. 
\begin{equation} \label{eq:propagation_v}
v_{propagation}^*=\dfrac{dx^*}{dt^*}
\end{equation}
The propagation velocity of the initial main structure and the two separated vortices can be evaluated from Figure~\ref{fig:vsbranch} from the slope of vortex trace; $v_a^*$ represents the propagation velocity before bursting, at which stage there is almost no difference between Reynolds numbers with only  $v_a^*$ of Re =2,000 slightly lower than others. After separation, the propagation velocity of both large vortex loop $v_{b2}^*$ and $v_{b1}^*$ are almost the same at $Re_\Gamma=20,000$ and $Re_\Gamma= 200,000$, and larger than those from the case $Re_\Gamma=2,000$. For a given Reynolds number, the small vortex loop propagates about three times faster than the large one. 

\begin{table} [h!]
\centering
\begin{tabular}{p{2cm}<{\centering} p{2cm}<{\centering} p{2cm}<{\centering} p{2cm}<{\centering}} 
\toprule
$Re_\Gamma$ & 2,000 & 20,000 & 200,000\\
\midrule                                                                                                                                                                                                                                                                                                                                                                                                                                                                                                                                                                                                                                                                                                                                                                                                                                                                                                                         
$v_a^*$ & 0.219 & 0.228 & 0.225 \\
$v_{b1}^*$ & 0.263 &  0.294 & 0.293 \\
$v_{b2}^*$ & 0.074 & 0.095 & 0.092 \\
\bottomrule 
\end{tabular}
\caption{Non-dimensional propagation velocity of the knotted vortex before burst, and that of separated vortex structures. }
\label{tab:velocity}
\end{table}

\newpage

\section{Conclusion}

In this work, the experimental setup for the study of knotted vortices has been reproduced and extented numerically by the means of state-of-the-art LES computations.
With a resolution of 10 grid points for the definition of vortex cores, the flow topology extracted from the simulation is found to be extremely similar to the experiment.
The numerical simulation of the trefoil knot vortex shape shows a vortex filament entanglement process in three locations of the main structure leading to the separation of this structure in two distinct vortices.
This entanglement process is found to be related with a strong increase of enstrophy in the flow.
This phenomenon is likely driven by small-scale dynamics and the sustained increase of enstrophy afterwards suggest the onset of background turbulence due to the entanglement process.
The examination of energy spectra show a sudden transfer of energy from large to small scales during the bursting process, characterized by the steepening of the spectra at high wavenumbers.
This effect is especially visible in the high Reynolds number computations.
The kinematics before breakdown is independent of Reynolds number and after that the separated vortex structures with higher Reynolds numbers are found to propagate faster than those at $Re_\Gamma = 2,000$.

\bibliographystyle{aiaa}
\bibliography{../../../biblio_latex/references}

\end{document}